\documentclass[proof]{WileyASNA-v1}
\usepackage{framed}
\usepackage{natbib}

\articletype{Original Article}

\received{-}
\revised{-}
\accepted{-}

\begin{document}

\title{Calculated brightness temperatures of solar structures compared with ALMA and Mets{\"a}hovi measurements}

\author[1]{Filip Matkovi\'c*}

\author[1]{Roman Braj\v{s}a}

\author[2,3]{Matej Kuhar}

\author[2,4]{Arnold O. Benz}

\author[5]{Hans -G. Ludwig}

\author[6,7]{Caius L. Selhorst}

\author[1]{Ivica Skoki\'{c}}

\author[1]{Davor Sudar}

\author[8]{Arnold Hanslmeier}

\authormark{Matkovi\'{c} \textsc{et al}}

\address[1]{\orgdiv{Hvar Observatory}, \orgname{Faculty of Geodesy, University of Zagreb}, \orgaddress{\state{Zagreb}, \country{Croatia}}}

\address[2]{\orgdiv{Institute for Data Science, University of Applied Sciences and Arts Northwestern Switzerland}, \orgaddress{\state{Windisch}, \country{Switzerland}}}

\address[3]{\orgdiv{Finstar, c/o Hypothekarbank Lenzburg AG}, \orgaddress{\state{Lenzburg}, \country{Switzerland}}}

\address[4]{\orgdiv{Institute for Particle Physics and Astrophysics, Department of Physics}, \orgaddress{\state{ETH Zurich}, \country{Switzerland}}}

\address[5]{\orgdiv{Landessternwarte K\"onigstuhl}, \orgname{Zentrum f\"ur Astronomie der Universit\"at Heidelberg}, \orgaddress{\state{Heidelberg}, \country{Germany}}}

\address[6]{\orgdiv{NAT - N\'ucleo de Astrof\'isica}, \orgname{Universidade Cidade de S\~ao Paulo}, \orgaddress{\state{S{\~a}o Paulo}, \country{Brazil}}}

\address[7]{\orgdiv{Center for Solar-Terrestrial Research}, \orgname{New Jersey Institute of Technology, Newark}, \orgaddress{\state{New Jersey}, \country{USA}}}

\address[8]{\orgdiv{Institute for Geophysics, Astrophysics and Meteorology, Institute of Physics}, \orgname{University of Graz}, \orgaddress{\state{Graz}, \country{Austria}}}

\corres{*Filip Matkovi\'{c}, Hvar Observatory, Faculty of Geodesy, University of Zagreb, Ka\v{c}i\'{c}eva 26, 10000 Zagreb, Croatia. \\\email{fmatkovic@geof.hr}}

\presentaddress{Croatian Science Foundation, project ID: 7549. Austrian-Croatian Bilateral Scientific Projects. Horizon 2020 project SOLARNET, project ID: 824135. Alexander von Humboldt Foundation. Deutsche Forschungsgemeinschaft, project ID: 138713538$-$SFB 881. S{\~a}o Paulo Research Foundation (FAPESP), grant number: 2019/03301-8.}


\jnlcitation{\cname{%
\author{Matkovi\'c F.}, 
\author{Braj{\v s}a R.}, 
\author{Kuhar M.}, 
\author{Benz A. O.},
\author{Ludwig H.-G.},
\author{Selhorst C. L.},
\author{Skoki\'c I.},
\author{Sudar D.}, and 
\author{Hanslmeier A.}} (\cyear{Year}), 
\ctitle{A comparison of the calculated and measured solar brightness temperatures in the mm and sub-mm wavelength ranges}, \cjournal{Astronomische Nachrichten}, \cvol{Vol. No.}.}

\abstract{The Atacama Large Millimeter/submillimeter Array (ALMA) allows for solar observations in the wavelength range of 0.3$-$10 mm, giving us a new view of the chromosphere. The measured brightness temperature at various frequencies can be fitted with theoretical models of density and temperature versus height. We use the available ALMA and Mets{\"a}hovi measurements of selected solar structures (quiet sun (QS), active regions (AR) devoid of sunspots, and coronal holes (CH)). The measured QS brightness temperature in the ALMA wavelength range agrees well with the predictions of the semiempirical Avrett$-$Tian$-$Landi$-$Curdt$-$W{\"u}lser (ATLCW) model, better than previous models such as the Avrett$-$Loeser (AL) or Fontenla$-$Avrett$-$Loeser model (FAL). We scaled the ATLCW model in density and temperature to fit the observations of the other structures. For ARs, the fitted models require 9\%$-$13\% higher electron densities and 9\%$-$10\% higher electron temperatures, consistent with expectations. The CH fitted models require electron densities 2\%$-$40\% lower than the QS level, while the predicted electron temperatures, although somewhat lower, do not deviate significantly from the QS model. Despite the limitations of the one-dimensional ATLCW model, we confirm that this model and its appropriate adaptations are sufficient for describing the basic physical properties of the solar structures.}


\maketitle

\section{Introduction}\label{Introduction}

The chromosphere is possibly the least observed and studied part of the solar atmosphere in the radio wavelength range. Therefore, it is crucial to study this layer of the atmosphere and to construct theoretical models that best describe the (radio) observations. An important opportunity to compare the results of theoretical models with measurements and to check their reliability is provided by the Atacama Large Millimeter/submillimeter Array (ALMA){\footnote{\url{http://www.almaobservatory.org}}$^{\rm ,}$\footnote{\url{https://www.eso.org/sci/facilities/alma.html}}}, which enable the observation of the solar chromosphere in the mm and sub-mm range (Bastian et al. \citeyear{Bastian2018}; Loukitcheva \citeyear{Loukitcheva2019}). ALMA observations provide a good basis for discrimination between various models and assessment of physical parameters of solar structures.

In the present analysis, we use the semiempirical modeling, in which a solar atmosphere model (electron density and temperature as a function of height) is constructed from the observed spectral lines of various elements, usually in the extreme-ultraviolet (EUV) and infrared parts of the spectrum. The model is used as is or modified for specific solar structures. The radiative transfer equation is then solved giving brightness temperature predictions. The first widely used atmospheric model of this type was the Vernazza$-$Avrett$-$Loeser (VAL) model (Vernazza et al. \citeyear{Vernazza1981}) and in each subsequent decade new and improved models have been published: the Fontenla$-$Avrett$-$Loeser (FAL) model (Fontenla et al. \citeyear{Fontenla1993}), the Avrett$-$Loeser (AL) model (Avrett \& Loeser \citeyear{Avrett2008}), and the Avrett$-$Tian$-$Landi$-$Curdt$-$W\"ulser (ATLCW) model (Avrett et al. \citeyear{Avrett2015}). In addition, a modification of this class of models was developed by Selhorst et al. (\citeyear{selhorst2005}) specifically to include spicules and to cover radio frequencies in the 2$-$400 GHz range (Selhorst$-$Silva$-$Costa (SSC) model).

In multiwavelength radio observations, the semiempirical models were used mostly to describe the quiet sun (QS). One such example is the earlier study by Benz et al. (\citeyear{Benz1997}), which used a modified and extended FAL model to interpret the radio emission from the quiet corona observed with the Very Large Array (VLA) at the wavelengths of 13 mm, 20 mm, and 36 mm. That study found that, at constant electron temperature in the FAL model, an rms fluctuation in density of 10.2\%, 17.6\%, and 16.4\% is required to produce the standard deviations in the QS brightness observed at the three wavelengths, respectively. Benz et al. (\citeyear{Benz1997}) also point out that the VAL model significantly exceeds the observed QS brightness temperature and deviates more from the observations than the FAL model.

The semiempirical QS models can also be used to study mm and sub-mm features of other solar structures. In previous studies, a similar modeling of the solar atmosphere using the FAL model was utilized to analyze the 8-mm radio emission from coronal holes (CHs), prominences (PRs) on the solar disc, and active regions (ARs) (Braj{\v s}a et al. \citeyear{Brajsa2007}, \citeyear{Brajsa2009}). Various models were constructed for a given solar structure by modifying the density and temperature of the FAL model, with thermal bremsstrahlung taken as the dominant radiation mechanism. An important result of those studies was that the assumed thermal bremsstrahlung explains well the observed radio emission of the given solar structures. The FAL model was further developed to calculate the brightness temperature of the QS, ARs, and CHs in the wavelength range of 0.3$-$10 mm (Braj{\v s}a et al. \citeyear{Brajsa2018a}). Various models were created and compared with the ALMA observations at the wavelength of 1.21 mm, and the properties (e.g., brightness, density, etc.) of the given structures were determined. In addition to previous studies, the SSC model was used to successfully interpret polar brightening observations at the wavelengths of 1.2, 3, and 18 mm on the Sun (Selhorst et al. \citeyear{selhorst2017}, \citeyear{Selhorst2019}).

The two newer AL and ATLCW models are based on similar or more accurate observations and better treatment of the structure of the solar atmosphere. The AL model is based primarily on atlases of the EUV solar spectrum from data collected by the Solar Ultraviolet Measurement of Emitted Radiation (SUMER) instrument aboard the Solar and Heliospheric Observatory (SOHO), as well as data from the High Resolution Telescope and Spectrograph (HRTS). The chromospheric part of the model is semiempirical with a temperature distribution adjusted to achieve optimal agreement between the calculated and observed continuum intensities, line intensities, and line profiles. The transition region model is theoretically determined from a balance between radiative losses and downward energy flow from the corona due to thermal conduction and particle diffusion. The boundary conditions at the base of the transition region were determined at the top of the chromosphere from the semi-empirical model.

On the other hand, the ATLCW model is an upgrade of the AL model with the main difference in the treatment of the transition region, where the temperature is adjusted to fit the H, He I, and He II observations and the slope of the Lyman continuum. The profile of the Mg II k line observed with the Interface Region Imaging Spectrograph (IRIS) was also calculated and taken into account when adjusting the calculated atmospheric temperature to the observations.

We now briefly describe the procedure performed in the present analysis. The starting point is the semiempirical ATLCW model of the QS atmosphere, giving density and temperature as a function of height. Such model is then used in its original form or modified for a specific solar structure, in our case AR and CH, as the input model. The modification is performed by multiplying the density and temperature by a numerical factor for the defined height range. Then the radiation mechanism is chosen, in this case thermal bremsstrahlung, and the radiative transfer equation is numerically solved for each input model. The output of the calculation is the radiation intensity (expressed by the brightness temperature) as a function of wavelength. This result is then compared with ALMA and other measurements, and the input model is modified again by changing the numerical factors, if necessary. The cycle is repeated until the best correspondence between calculated and measured brightness temperatures is reached. The final results are the best-fitting model and the goodness of the model fit. The main idea of this work is to compare the predictions of AL and ATLCW models with QS measurements from ALMA and Mets\"ahovi observatories, using thermal bremsstrahlung as the dominant radiation mechanism and improved values of the Gaunt factor. Then, the better model (in our case ATLCW model) is modified to fit the observations of three different solar regions (QS, AR, and CH) returning their estimated electron temperatures and densities.

\section{The models and the brightness temperature calculation}\label{Calculation}

The main radiation mechanism of the quiet and active chromosphere at mm and sub-mm wavelengths is thermal bremsstrahlung (Zirin \citeyear{Zirin1988}; Hurford \citeyear{Hurford1992};  Braj{\v s}a \citeyear{Brajsa1993}; White \citeyear{White2002}; Benz \citeyear{Benz2009}; Wedemeyer et al. \citeyear{Wedemeyer2016}) and it is assumed here for interpretation and calculations. In radio astronomy, the measured radiation intensities are usually expressed in terms of the brightness temperature. Following, for example, Wilson et al. (\citeyear{Wilson2013}), the brightness temperature can be derived starting from the radiation transfer equation:
\begin{equation}
\label{Eq_radiative_transfer1}
I_\lambda=\int_{0}^{\infty}B_\lambda e^{-\tau_\lambda}{\rm d}\tau_\lambda,
\end{equation}
where $\lambda$ is the measuring wavelength, $\tau_\lambda$ is the optical depth, and $I_\lambda$ is the measured radiation intensity. If we assume that the source function $B_\lambda$ in Equation (\ref{Eq_radiative_transfer1}) follows Planck's law and we take the Rayleigh$-$Jeans approximation ($hc/\lambda\ll k_\mathrm{B}T_\mathrm{e}$) into account, we can write the source function in the form:
\begin{equation}
\label{Planck}
B_\lambda=\frac{2hc^2}{\lambda^5\left(e^{\frac{hc}{\lambda k_\mathrm{B}T_\mathrm{e}}}-1\right)}\approx\frac{2ck_\mathrm{B}T_\mathrm{e}}{\lambda^4},
\end{equation}
where $h$ is the Planck's constant, $k_\mathrm{B}$ is the Boltzmann's constant, $c$ is the speed of light, and $T_\mathrm{e}$ is the electron temperature. If we substitute the $B_\lambda$ function in Equation (\ref{Eq_radiative_transfer1}) with the last expression in Equation (\ref{Planck}) and take into account that the brightness temperature is the solution of the radiative transfer equation, we now write Equation (\ref{Eq_radiative_transfer1}) in a new form (e.g., Braj{\v s}a et al. \citeyear{Brajsa2018a}):
\begin{equation}
\label{Eq_radiative_transfer}
T_\mathrm{b}(\lambda)=\int_{0}^{\infty}T_\mathrm{e}e^{-\tau_\lambda}{\rm d}\tau_\lambda,
\end{equation}
where $T_\mathrm{b}(\lambda)=\lambda^4I_\lambda/(2ck_\mathrm{B})$ is the brightness temperature. In the present work, we use Equation (\ref{Eq_radiative_transfer}) when calculating the brightness temperature as a function of wavelength. As the wavelength decreases, $\tau_\lambda$ reaches $\tau_\lambda=1$ at lower heights in the solar atmosphere where the temperature is lower. The optical depth d$\tau_\nu$ for the observing frequency $\nu$ of bremsstrahlung for solar abundances and distance d$s$ neglecting the magnetic field is given by (e.g., Benz \citeyear{Benz2002}):
\begin{equation}
\label{Eq_optical_depth}
{\rm d}\tau_\nu=\frac{0.01146\;\mathrm{cm^{5}Hz^{2}K^{3/2}}\times\ln\Lambda n^{2}_{e}}{\left(1-8.06\times 10^{7}\;\mathrm{cm^3Hz^{2}}\times n_{e}/\nu^{2}\right)^{1/2}\nu^{2}T_{e}^{3/2}} {\rm d} s  .
\end{equation}
The Gaunt factor, $\ln\Lambda$, is a slowly varying function of electron density, $n_{e}$, and temperature, $T_e$ (Rybicki \& Lightman \citeyear{Rybicki1985}; Weinberg \citeyear{Weinberg2020}). In the present work, the Gaunt factor for the observed frequency is calculated using the accurate interpolation method recently developed by van Hoof et al. (\citeyear{vanHoof2014}) and implemented in the solar context by Sim\~oes et al. (\citeyear{Simoes2017}) and Selhorst et al. (\citeyear{Selhorst2019}).

\subsection{Obtaining theoretical models}

As a reference model of the solar atmosphere, we use the ATLCW QS model in its original form. To obtain the brightness temperature profile, we calculate the brightness temperature for a given wavelength using the input $T_{e}$ and $n_{e}$ values from the ATLCW model and store its increase per unit height in an array. These contributions are integrated over all heights defined for a given solar structure, yielding the total brightness temperature for a given wavelength (Equation \ref{Eq_radiative_transfer}). This procedure is repeated for all given wavelengths. We note that the calculation procedure in the present work is similar to that used by Braj{\v s}a et al. (\citeyear{Brajsa2018a}), but with important differences. Braj{\v s}a et al. (\citeyear{Brajsa2018a}) used the FAL model (Fontenla et al. \citeyear{Fontenla1993}) and the Gaunt factor calculation according to Bekefi (\citeyear{Bekefi1966}) and Benz (\citeyear{Benz2002}) and here we use the ATLCW model and the Gaunt factor according to van Hoof et al. (\citeyear{vanHoof2014}) and Selhorst et al. (\citeyear{Selhorst2019}).

In order to use the ATLCW QS model as input for the brightness temperature calculation of a given solar structure, it must be modified by scaling the $n_\mathrm{e}$ and $T_\mathrm{e}$ values of the model with multiplicative factors, $f_\mathrm{n}$ and $f_\mathrm{T}$, defined as:
\begin{equation}
\label{multi_factors}
f_\mathrm{n}=\frac{n_\mathrm{e}\mathrm{(modified)}}{n_\mathrm{e}\mathrm{(original)}},\; f_\mathrm{T}=\frac{T_\mathrm{e}\mathrm{(modified)}}{T_\mathrm{e}\mathrm{(original)}},
\end{equation}
where $n_\mathrm{e}$ and $T_\mathrm{e}$ denote electron densities and temperatures of the original and modified models. The model is modified only in the height range defined for a given solar structure. For ARs, we assumed a minimum height of 400 km and a maximum of 57\;797 km (maximum height value available in the ATLCW QS model). For CHs, we assumed the height range of 2\;100$-$57\;797 km, where the minimum height of 2\;100 km was chosen because at about this height the transition region for the used QS model ($T = 2.4 \times 10^5$ K) ends. For QS on the other hand, the entire height range from the photosphere to the maximum available height in the ATLCW QS model was considered.

\subsection{Fitting theoretical models to observational data}

In order to compare the brightness temperature from the theoretical model and actual measurements, we use the standard $\chi^2$-minimization technique (e.g., Ivezi\'c et al. \citeyear{Ivezic2014}) to fit the theoretical model on the measured data. The resulting brightness temperature profile obtained from the calculation procedure described above is compared with the observed values and a dimensionless $\chi^2$ analysis is performed, where we calculate the sum of $\chi^2$ values for all the observed wavelengths $\lambda$ using the following form:
\begin{equation}
\label{chi_squared}
\chi^2=\sum_\lambda{\left(\frac{x_{\mathrm{obs}} - x_{\mathrm{exp}}}{\Delta x_{\mathrm{obs}}}\right)^2},
\end{equation}
where $x_{\mathrm{obs}}$ and $x_{\mathrm{exp}}$ correspond to the observed brightness temperature and the prediction of a theoretical model, respectively, and $\Delta x_{\mathrm{obs}}$ is the measurement uncertainty of the observed brightness temperature. The best-fitting brightness temperature profile minimizes the above sum. This is done by varying values of the multiplicative factors $f_\mathrm{n}$ and $f_\mathrm{T}$ (Equation \ref{multi_factors}) and for each pair of the two multiplicative factors the brightness temperature calculation is repeated until the global minimum of Equation (\ref{chi_squared}) is found. The range of values within which the two multiplicative factors are varied is set so that the input model's combination of $n_\mathrm{e}$ and $T_\mathrm{e}$ matches observations within the corresponding height range. As a result, we obtain the best-fitting brightness temperature profile with the corresponding $f_\mathrm{n}$ and $f_\mathrm{T}$ factors. The uncertainties of the  $f_\mathrm{n}$ and $f_\mathrm{T}$ factors are calculated from the $\chi^2$ function where the $\chi^2$ value reaches $\Delta\chi^2=1$ above the minimum ($\chi^2_\mathrm{min}$), which corresponds to 1$\sigma$ uncertainty.

\section{Observational results}
\label{observational_results}

\subsection{ALMA data}

ALMA provides single-dish (White et al. \citeyear{White2017}) and interferometric (Shimojo et al. \citeyear{Shimojo2017}) observations of the full solar disc and a smaller area on the Sun, respectively. The wavelength bands most frequently used by ALMA are Band 3 and 6, which are centered around the wavelengths of 3 mm and 1.21 mm, respectively. These two bands cover the solar atmosphere in the height range of 600$-$1\;600 km for Band 3 and 400$-$1\;400 km for Band 6, with the highest sensitivity at around 960 km for Band 3 and at around 730 km for Band 6 (see Figure 5 in Wedemeyer et al. \citeyear{Wedemeyer2016}). In Braj{\v s}a et al. (\citeyear{Brajsa2018b}), the results of the first analysis of solar structures in the Band 6 full-disc ALMA image ($\lambda=1.21$ mm, beam size = 26 arcsec) are presented. The measurements of the brightness temperature of various structures (QS, AR, and CH) for December 18, 2015 are given in Table 1 in that paper. These results are repeated here in Table \ref{Table_2} in a slightly modified form.
\begin{table*}[h!]
\centering
\caption{~Measurements of brightness temperature $T_\mathrm{b}$(Structure) of various structures in solar atmosphere (central QS, AR (devoid of sunspots), and CH region) obtained by ALMA and Mets\"ahovi at corresponding time (Date) and wavelength or frequency ($\lambda / \nu$), with a given spatial resolution (Beam size). The brightness temperature $T_\mathrm{b}$(QS) corresponds to the brightness temperature measurement of a QS region at a similar distance from the solar disk's center as the given structure (central QS, AR, or CH). In the case of Mets\"ahovi, due to limb brightening already being taken into account, all its measured $T_\mathrm{b}$(QS) values correspond to that of the central QS region. The brightness temperature difference $\Delta T_\mathrm{b}$ calculated as $\Delta T_\mathrm{b}=T_\mathrm{b}$(Structure) $-$ $T_\mathrm{b}$(QS) is also given. For details on the individual measurement see the corresponding references (Reference).}
\label{Table_2} 
\centering
\resizebox{2\columnwidth}{!}{
\begin{tabular}{c c c c c c c c c}
\hline\midrule
Structure & Instrument & Date & $\lambda$/$\nu$ & Beam size & $T_\mathrm{b}$(QS) & $T_\mathrm{b}$(Structure) & $\Delta T_\mathrm{b}$ & Reference \\
(name) & (name) & (y:m:d) & (mm)/(GHz) & (arcsec) & (K) & (K) &  (K) & (citation)\\
\midrule
\multirow{10}{*}{QS} &   ALMA   &   2020-08-01   & 0.86/347 & 21 & 5985 & 5985 &   0   & Alissandrakis et al. (\citeyear{Alissandrakis2023}) \\ 
             &   ALMA    &  2015-12-18  &  1.21/248 & 26 &6040 &  6040 &     0   & Braj{\v s}a et al. (\citeyear{Brajsa2018b})  \\ 
             &   ALMA  &   2020-01-04   &   1.50/198 & $29$ & 6467 & 6467 &    0    &  Alissandrakis et al. (\citeyear{Alissandrakis2023})\\
&   ALMA    &  2015-12-17    & 2.80/107 & 58 & 7114 & 7114 &  $0$       &  Present work \\ 
             &   \multirow{2}{*}{Mets\"ahovi}  &  1994-10-18 (start)    &  \multirow{2}{*}{3.40/87} & \multirow{2}{*}{60} & \multirow{2}{*}{7200} & \multirow{2}{*}{7200} &      \multirow{2}{*}{0}     & \multirow{2}{*}{Urpo et al. (\citeyear{Urpo1997})}     \\
	     &   &  1995-10-15 (end)    &      &&  &  &  &    \\
             &   \multirow{2}{*}{Mets\"ahovi}  &   1994-10-18 (start)  &  \multirow{2}{*}{3.90/77} & \multirow{2}{*}{72} & \multirow{2}{*}{7250} & \multirow{2}{*}{7250} &       \multirow{2}{*}{0}     &    \multirow{2}{*}{Urpo et al. (\citeyear{Urpo1997})}   \\
	     &   &  1995-10-15 (end)    &      &&  &  &  &    \\
             &  \multirow{2}{*}{Mets\"ahovi}  & 2018-03-17 (start) & \multirow{2}{*}{8.10/37} & \multirow{2}{*}{144} & \multirow{2}{*}{8100} & \multirow{2}{*}{8100} &   \multirow{2}{*}{0}    &  \multirow{2}{*}{Kallunki \& Tornikoski (\citeyear{Kallunki2018})} \\
	     &   &  2018-05-16 (end)    &      &&  &  &  &    \\
\midrule
       \multirow{3}{*}{AR}      &   ALMA    &  2015-12-18    &  1.21/248 & 26 & 6240 & 7250 &   $+1010$       &   Braj{\v s}a et al. (\citeyear{Brajsa2018b}) \\ 
&   ALMA   &  2015-12-17   &  2.80/107 & 58 & 7303 & 7928&  $+625$       &   Present work \\ 
             &   Mets\"ahovi  &   2015-12-19   &  8.10/37 & 144 & 8100& 8580 &  $+480$         &   Present work  \\
\midrule
     \multirow{3}{*}{CH}        &   ALMA    &   2015-12-18   & 1.21/248 & 26 & 6590 & 6540 & $-50$       &   Braj{\v s}a et al. (\citeyear{Brajsa2018b}) \\ 
&   ALMA    &  2015-12-17    &  2.80/107 & 58 & 7543 & 7398 &  $-145$       &   Present work \\ 
             &   Mets\"ahovi  &   2015-12-19  &  8.10/37 &144 & 8100  & 8020 & $-80$  &Present work  \\
\midrule
\end{tabular}
}
\end{table*}

In the present work, we extend the analysis of Braj{\v s}a et al. (\citeyear{Brajsa2018b}) by including an analysis of the same solar structures in the Band 3 ALMA full-disc map ($\lambda=2.80$ mm, beam size = 58 arcsec) taken on the previous day, December 17, 2015. Following the procedure described in Braj{\v s}a et al. (\citeyear{Brajsa2018b}), we determine the brightness temperature of the QS region in the center of the solar disc for the Band 3 data by averaging the values of all pixels within a radius of about 15 pixels. The size of a single pixel of the Band 3 full-disc map is 6 arcsec on the Sun. For CH, which is a southern polar CH, and thus at the solar limb, the brightness temperature is measured by averaging the value within an area of a radius of 10 pixels within the visible CH structure, but away from the solar limb. Next, the entire AR, which is observed near the central region of the solar disk, is selected based on a numerical criterion by setting the lower threshold for the brightness temperature. The brightness temperature is then averaged over all pixels brighter than this threshold. In this way, sunspots are automatically excluded from the AR and do not contribute to the measurement of the AR brightness temperature.

Limb brightening was taken into account by comparing the brightness temperature measurements of each non-QS structure with the brightness temperature values averaged over a 10-pixel radius QS region located at the same radial distance from the solar center as the observed non-QS structure. Unlike earlier studies by Alissandrakis et al. (\citeyear{Alissandrakis2017}), Selhorst et al. (\citeyear{Selhorst2019}), and Sudar et al. (\citeyear{Sudar2019}), we applied the limb brightening correction to the measured brightness temperatures before the modeling was done. The measurements of the brightness temperature are listed in Table \ref{Table_2}.

In addition to the ALMA Band 3 and Band 6 measurements, we also included the observational results of Alissandrakis et al. (\citeyear{Alissandrakis2022}) for the Band 7 mode ($\lambda=0.86$ mm, beam size = 21 arcsec) and an estimate for the Band 5 mode ($\lambda=1.50$ mm, beam size = 29 arcsec) also from that paper. Here we use the brightness temperature values from the corrigendum Alissandrakis et al. (\citeyear{Alissandrakis2023}), in which, unlike the original paper Alissandrakis et al. (\citeyear{Alissandrakis2022}), the frequency variation of the Gaunt factor was taken into account. The results for Band 5 and 7, available only for QS, are shown in Table \ref{Table_2}.

\subsection{Mets\"{a}hovi data}
Due to the current unavailability of the ALMA measurements above the wavelength of 3 mm, we are missing a large part of the mm range reachable for ALMA, which could be important for theoretical modeling of the observed solar features. For this reason, we extend the currently observed ALMA wavelength range to longer wavelengths by including observations from the Mets\"ahovi 14-m diameter radio telescope\footnote{\url{https://www.aalto.fi/en/metsahovi-radio-observatory}}. The Cassegrain telescope system of the Mets\"ahovi radio telescope can be used to obtain full-disc and partial maps of the Sun in the wavelength range of 3 mm$-$3 cm (Urpo et al. \citeyear{Urpo1997}).

In this work, we use full-disc Mets\"ahovi observations of the mentioned solar structures at 3.40 mm (beam size = 60 arcsec), 3.90 mm (beam size = 72 arcsec), and 8.10 mm (beam size = 144 arcsec). The Mets\"ahovi results for the brightness temperature were obtained using a similar procedure as described for ALMA, but with the limb brightening effect already accounted for, are shown in Table \ref{Table_2}.

\section{Modeling results and comparison with observations}
\label{Results}

Following the procedure described in Section \ref{Calculation}, we obtained the best-fitting brightness temperature profiles using the modified ATLCW QS atmosphere model as the input model in our calculations. In the following subsections, we present the modeling results separately for three different solar structures: QS region in the center of the solar disc, AR devoid of sunspots, and a CH region.

\subsection{Quiet sun}
\label{quiet_sun}

\begin{figure}[h!]
\centering
\resizebox{0.97\hsize}{!}{\includegraphics{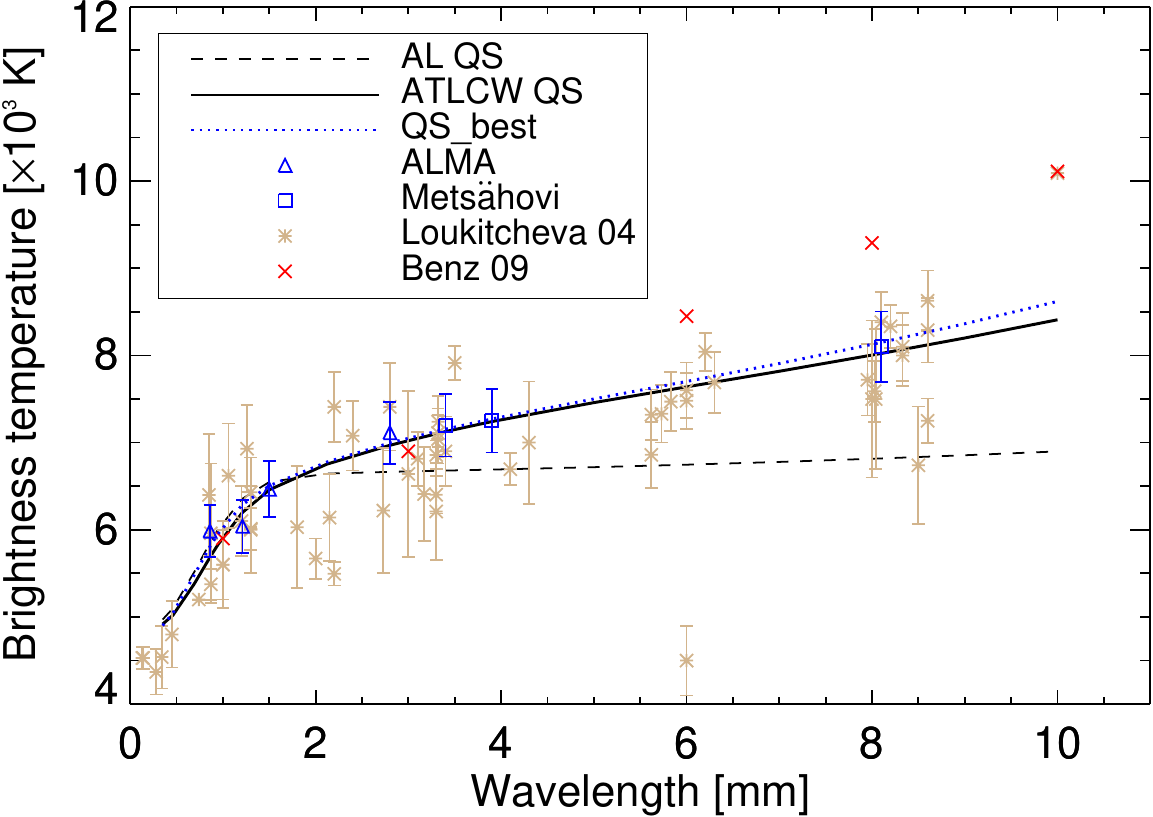}}
\caption{~Calculated QS brightness temperature profiles. The black curves indicate the brightness temperature profiles obtained using the original AL QS (dashed curve) and ATLCW QS models (solid curve), while the blue dotted curve indicates the best-fitting QS profile, obtained using the modified ATLCW QS model, for ALMA and Mets\"ahovi measurements (blue symbols). For comparison, the QS measurements compiled by Loukitcheva et al. (\citeyear{Loukitcheva2004}) (beige symbols) and Benz (\citeyear{Benz2009}) (red symbols) are also indicated.}
\label{Fig_1}
\end{figure}
The first solar structure is the quiet sun (QS), which is regarded as the region with a diffuse emission devoid of highly active structures (e.g., AR). Due to being the largest and widespread solar structure with similar properties throughout it, for measurements and modeling analysis of the brightness temperature, for the QS structure only a small part of the QS region in the center of the solar disk is considered (see Section \ref{observational_results}).

The calculated best-fitting QS brightness temperature profile (blue dotted curve) obtained for the ALMA and Mets\"ahovi measurements of the central QS structure in Table \ref{Table_2} using the modified ATLCW QS model is shown for the ALMA wavelength range in Figure~\ref{Fig_1}. We note that the error bars for the ALMA and Mets\"ahovi data in all figures refer to measurement errors and not the distribution of the mean values of all data points. These measurement errors are estimated to be about 5\%$-$10\% (Shimojo et al. \citeyear{Shimojo2017}; White et al. \citeyear{White2017}). Other earlier and much older observational results for the QS compiled by Loukitcheva et al. (\citeyear{Loukitcheva2004}) and by Benz (\citeyear{Benz2009}), which have low spatial, spectral, and temporal resolutions, are added here for comparison only.

We see that the original ATLCW QS model (black solid curve) reproduces well the ALMA and Mets\"ahovi observations over the entire ALMA wavelength range. Even the best-fitting QS model does not differ significantly from the original one (to be distinguished from the modified ones), confirming the accuracy of the ATLCW QS model and justifying our use of it as a reference model. In comparison, the profile resulting from the older original AL QS model (black dashed curve) differs significantly from both the measurements and the profile from the original ATLCW QS model already after a wavelength of 2 mm. Moreover, the observations reviewed by Loukitcheva et al. (\citeyear{Loukitcheva2004}) and by Benz (\citeyear{Benz2009}) also appear to be in better agreement with the original ATLCW model than with the original AL model. Here the measurements compiled by Loukitcheva et al. (\citeyear{Loukitcheva2004}) follow both the original and modified ATLCW models well over the entire ALMA wavelength range, while those of Benz (\citeyear{Benz2009}) agree well only up to a wavelength of 6 mm and begin to depart upward from the ATLCW models at longer wavelengths.

The density and temperature multiplicative factors, $f_\mathrm{n}$ and $f_\mathrm{T}$ respectively (QS\_best in Table \ref{Table_3}), correspond to the best fit of the QS brightness temperature profile. The resulting value of the $f_\mathrm{T}$ factor differs from unity by only 1\%, which shows an excellent prediction of the original ATLCW model for the electron temperature of the QS. On the other hand, the resulting $f_\mathrm{n}$ factor shows that the QS density is 13\% higher than the value predicted by the original ATLCW model. However, this small deviation in density did not result in any significant change in the best-fitting brightness temperature profile within the ALMA wavelength range, indicating the validity of the original ATLCW QS model.
\begin{table*}[h!]
\centering
\caption{~Output multiplicative factors for electron density ($f_\mathrm{n}$ best fit) and temperature ($f_\mathrm{T}$ best fit) corresponding to the best-fitting model for two measurement procedures (index a and b) calculated for various solar structures (central QS, AR (devoid of sunspots), and CH region) with corresponding minimum $\chi^2_\mathrm{min}$ value determined from Equation (\ref{chi_squared}). The $f_\mathrm{n}$ range and $f_\mathrm{T}$ range correspond to the range of density and temperature factor values (Equation \ref{multi_factors}), within which the values of the two factors are varied to modify the density and temperature parameters of the input atmospheric model globally within a given height range until the global minimum of Equation (\ref{chi_squared}) is found, thus producing the best fit. For a detailed description of the fitting procedure and the factor calculation, see Section \ref{Calculation}.}
\label{Table_3} 
\centering
\resizebox{2\columnwidth}{!}{
\begin{tabular}{c c c c c c c c c c}
\hline\midrule
Structure& Input atm. model & Height range & $f_\mathrm{n}$ range & $f_\mathrm{T}$ range &Procedure& $f_\mathrm{n}$ best fit & $f_\mathrm{T}$ best fit  & $\chi^2_{\mathrm{min}}$ & Ref. name\\
(name)& (name) & (km) & (value) & (value) & (index)& (value) &(value)  &(value) & (name)\\
\midrule
\multirow{1.2}{*}{QS} & \multirow{1.2}{*}{ATLCW QS} &   \multirow{1.2}{*}{$0-57\;797$}   &\multirow{1.2}{*}{$0.5-1.5$} & \multirow{1.2}{*}{$0.5-1.5$}  & a& $1.13^{+0.14}_{-0.13}$ & $0.99^{+0.02}_{-0.02}$ & $1.19$ &\multirow{1.2}{*}{QS\_best}\\
\midrule
\multirow{2.5}{*}{AR}  &\multirow{2.5}{*}{ATLCW QS}&   \multirow{2.5}{*}{$400-57\;797$}    & \multirow{2.5}{*}{$1-3$} & \multirow{2.5}{*}{$1-3$}  &  a& $1.09^{+0.18}_{-0.16}$ & $1.10^{+0.04}_{-0.03}$ & $0.76$  &\multirow{2.5}{*}{AR\_best}\\ [0.5em]
  &&&  && b& $1.13^{+0.17}_{-0.16}$ & $1.09^{+0.03}_{-0.04}$  & $1.71$ & \\ 
\midrule
\multirow{8.5}{*}{CH}  &   \multirow{8.5}{*}{ATLCW QS}    &  \multirow{8.5}{*}{$2\;100-57\;797$}  &\multirow{2.9}{*}{$1/5-1/0.5$} &\multirow{2.9}{*}{$1/25-1/0.5$}&a& $1/\left(1.34^{+1.34}_{-0.33}\right)$& $1/\left(5.40^{+>20}_{-5.00}\right)$ & $0.49$& \multirow{2.9}{*}{CH1\_best}\\[0.5em] 
 &&&&&b& $1/\left(1.67^{+3.84}_{-0.75}\right)$& $1/\left(5.76^{+>20}_{-5.66}\right)$ &  $0.18$  &\\ \cmidrule{4-10}
  &&&\multirow{2.5}{*}{$1/5-1/0.5$}&\multirow{2.5}{*}{$1/5-1/0.5$}&a& $1/\left(1.08^{+1.07}_{-0.40}\right)$& $1/\left(1.36^{+>4}_{-1.22}\right)$ &  $0.49$  &\multirow{2.5}{*}{CH2\_best}\\[0.5em]
  &&&&&b& $1/\left(1.33^{+3.08}_{-0.63}\right)$& $1/\left(1.35^{+>4}_{-1.24}\right)$ & $0.18$ & \\ \cmidrule{4-10}
  &&&\multirow{2.5}{*}{$1/5-1/0.5$}&\multirow{2.5}{*}{$1$}&a& $1/\left(1.02^{+1.01}_{-0.25}\right)$&\multirow{2.75}{*}{$1$}& $0.49$  &\multirow{2.5}{*}{CH3\_best}\\[0.5em] 
  &&&&&b&$1/\left(1.26^{+1.23}_{-0.56}\right)$&& $0.18$ & \\ 
\midrule
\end{tabular}
}
\end{table*}

\subsection{Active region}
\label{active_regions}

The second structure is the active region (AR), and it is the region with the highest activity on the Sun with a very enhanced radiation emission in comparison with other regions. In this work, sunspots present inside an AR are considered as a separate structure and are excluded from the AR as described in Section \ref{observational_results}. However, unlike the QS, the entire AR structure, with exclusion of sunspots, is considered in our analysis.

\begin{figure}[h!]
\centering
\resizebox{0.97\hsize}{!}{\includegraphics{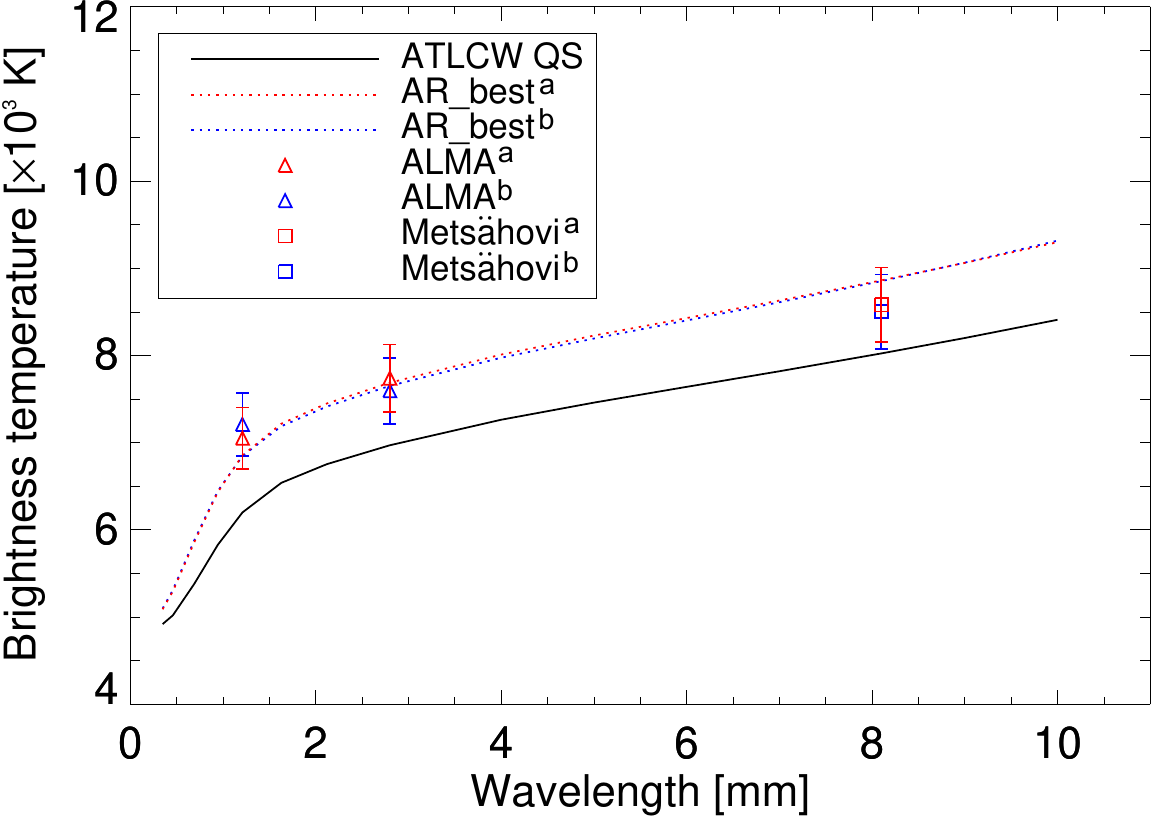}}
\caption{~Similar to Figure~\ref{Fig_1} for AR. The red and blue dotted curves indicate the best-fitting AR brightness temperature profile for ALMA and Mets\"ahovi measurements obtained using the first (index a) and second measurement procedure (index b), respectively.}
\label{Fig_2}
\end{figure}

The modeling of ARs was similar to that of QS presented in the previous section, but compared with QS, the ALMA and Mets\"ahovi brightness temperature of ARs is determined using two measurement procedures. In the first procedure, which corresponds to the actual measurement, the brightness temperature of AR is determined by adding the difference $\Delta T_\mathrm{b}$ from Table~\ref{Table_2} to the measured brightness temperature of the central QS (QS structure in Table \ref{Table_2}), while in the second procedure, the difference $\Delta T_\mathrm{b}$ is added to the value given by the profile from the original ATLCW QS model (black curve in Figure~\ref{Fig_2}) for the corresponding wavelength. The results for both procedures are indicated by index a and b, respectively, in Figure~\ref{Fig_2} and Table \ref{Table_3}. This approach was chosen to investigate whether a small change in the measurements would lead to a significant change in the output brightness temperature profile. A similar approach was also used for CH (Figure \ref{Fig_5}).

The best-fitting AR profiles in Figure \ref{Fig_2} fit the ALMA and Mets\"ahovi measurements well, with no significant differences between the profiles of the two types of measurement a and b over the entire ALMA wavelength range. We should note that the Mets\"ahovi measurement is somewhat lower than predicted by, but still within errors from the AR profile. Moreover, both the observations and the AR profiles lie significantly higher than the brightness temperature profile for the original ATLCW QS model, indicating higher temperatures than the QS values. Assuming thermal bremsstrahlung as the main radiation mechanism higher brightness temperatures in ARs are mainly a consequence of enhanced density in the chromosphere and corona. This shifts the $\tau = 1$ layer to higher altitudes where the temperature is higher, as shown in Figure \ref{Fig_2}.

Both measurement procedures also resulted in very similar $f_\mathrm{n}$ and $f_\mathrm{T}$ factors (AR\_best in Table \ref{Table_3}), as expected from the appearance of the output AR profiles. Based on the resulting $f_\mathrm{n}^{\mathrm{a,b}}$ best fit factors, the density of the AR is 9\%$-$13\% higher than the QS level. This is consistent with the previously mentioned higher brightness of the ARs due to the enhanced plasma density of the ARs compared to the surrounding regions. In the case of $f_\mathrm{T}^{\mathrm{a,b}}$ best fit factors, the results show that the temperature of AR is 9\%$-$10\% higher than that of QS.

\subsection{Coronal hole}

Finally, we come to coronal hole (CH), a region of lower temperature and density in the solar atmosphere, which gives it a darker appearance in comparison to the surrounding regions. The CH in question was a southern polar CH (see Figure 1 in Braj{\v s}a et al. \citeyear{Brajsa2018b}), whose entire structure was not visible, so only a part of the CH structure away from the solar limb was considered for the CH brightness temperature analysis, as described in Section \ref{observational_results}.

Given the observed lack of temperature enhancements in the chromosphere and transition region, it is sufficient that the atmosphere models differ only in density at the transition region and coronal heights. As the coronal radiation is optically thin, a change in coronal temperature would not result in a significant change in brightness temperature. Here we investigate three cases, in which the same $f_\mathrm{n}$ factor range is chosen based on the hybrid network model of Gabriel (\citeyear{Gabriel1992}), with modification and upgrade taking into account, and its various studies of the solar corona (Stix \citeyear{Stix1989}; Hara et al. \citeyear{Hara1994}; Doschek et al. \citeyear{Doschek1997}; Golub \& Pasachoff \citeyear{Golub1997}; Gallagher et al. \citeyear{Gallagher1999}; Avrett \citeyear{Avrett2000}; Koutchmy \citeyear{Koutchmy2000}; Lang \citeyear{Lang2000}; Aschwanden \citeyear{Aschwanden2004}). For the first case, however, a wide range of $f_\mathrm{T}$ factors is chosen with no a priori assumption, but in two other cases, we restrict that range to values closer to 1 (second case) or fix the factor value to exactly 1 (third case).
\begin{figure}[h!]
\centering
\resizebox{0.97\hsize}{!}{\includegraphics{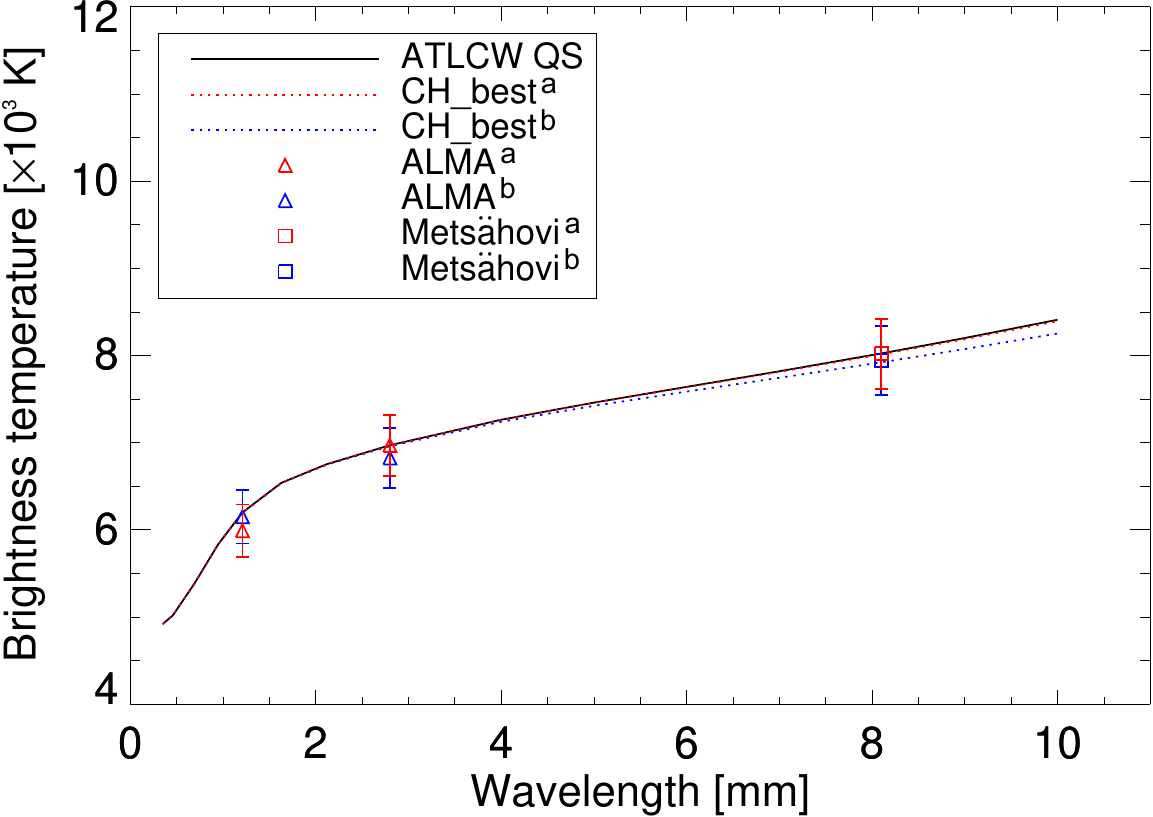}}
\caption{~Similar to Figure~\ref{Fig_2} for CH. The same figure is obtained for all three solutions of the density and temperature multiplicative factors, $f^{\mathrm{a,b}}_\mathrm{n}$ best fit and $f^{\mathrm{a,b}}_\mathrm{T}$ best fit respectively, denoted as CH1\_best, CH2\_best, and CH3\_best in Table \ref{Table_3}.}
\label{Fig_5}
\end{figure}

All three cases (CH1\_best, CH2\_best, and CH3\_best in Table \ref{Table_3}) resulted in similar CH brightness temperature profiles, as shown in Figure \ref{Fig_5}. Here, the CH profiles follow all measurements well. The measurements lie nearly at or slightly below the QS level from the original ATLCW model (black solid curve), indicating that CH is slightly darker than the surrounding QS atmosphere in the ALMA wavelength range. In addition, the measurement Procedures a and b show no significant difference in the output CH profile. However, the profile for Procedure a is essentially at the QS level, indicating no difference in the brightness temperature between CH and QS, while for Procedure b there is a small but increasing difference between CH and QS toward longer wavelengths.

Moreover, the fact that all three cases of $f^{\mathrm{a,b}}_\mathrm{T}$ best fit factors (Table \ref{Table_3}) give the same CH profile and that the uncertainty of those factors is higher than the actual factor value, but decreasing as it approaches 1, indicates that the CH brightness temperature does not depend on the electron temperature. If we would set the measurement uncertainties to lower values than they are given in this work, the density factor uncertainty becomes much smaller, but the uncertainty of the temperature factor still remains significantly high and going outside the considered value range. Therefore, a significant change in electron temperature does not result in a significant change in the CH brightness temperature, as is predicted. One can then simply set the temperature factor to 1 like in the third case. Between cases, the $f^{\mathrm{a,b}}_\mathrm{n}$ best fit factors (Table \ref{Table_3}) do not change significantly for both Procedures a and b. Procedure b yields a slightly lower density than Procedure a. However, the same resulting $f_\mathrm{n}$ and $f_\mathrm{T}$ factors of the best fit show that CH is less dense than QS, but also slightly colder, which is in good agreement with the expected appearance of CHs in the solar atmosphere (Baranovskii et al. \citeyear{Baranovskii2019}; Gopasyuk et al. \citeyear{Gopasyuk2020}).

\section{Discussion and conclusions}
\label{discussion_conclusion}

A comparison of the QS brightness temperature profiles calculated from the ATLCW QS atmosphere model with the recent ALMA and Mets\"ahovi observations (Figure~\ref{Fig_1}) reveals general agreement over the entire ALMA wavelength range. The best-fitting QS model showed no significant deviation from the original ATLCW model in both density and temperature. These results show that the original ATLCW QS model is sufficient for describing the QS atmosphere and no additional modifications are needed in the model at this time.

In the ALMA and Mets\"ahovi full-disc images, the ARs appear significantly brighter than the measured surrounding QS atmosphere, but also brighter than the ATLCW QS level (Figure~\ref{Fig_2}). At mm and sub-mm wavelengths, ARs are bright primarily due to increased density, and our AR models appear to be in a good agreement with this statement. We note that the electron temperature was also higher for ARs. The excess temperature of ARs compared to QS is consistent with the expected AR features where a strong magnetic field could lead to an increase in temperature, for example, due to dissipation in current sheets (da Silva Santos et al. \citeyear{daSilvaSantos2022}).

Using the SSC atmosphere model for the QS, de Oliveira e Silva et al. (\citeyear{deOliveiraeSilva2022}) modeled the density and temperature for different parts of an AR (umbra, penumbra, and plage) up to 6~000 km above the photosphere. The AR in question is the same as the one observed in the present work and in Braj{\v s}a et al. (\citeyear{Brajsa2018b}). Devoid of sunspot structures (umbra and penumbra), in the chromosphere, de Oliveira e Silva et al. (\citeyear{deOliveiraeSilva2022}) obtained the average AR density about 47\% higher and a temperature about 7\% higher than the SSC QS values. Our results for the AR $f_\mathrm{T}$ best fit factor (Table \ref{Table_3}) somewhat agree with the results by de Oliveira e Silva et al. (\citeyear{deOliveiraeSilva2022}), while our $f_\mathrm{n}$ best fit factors (Table \ref{Table_3}) seem to be in the lower range of their model. However, their value of 47\% higher density would still produce an AR profile within the measurement errors and also close to the AR profiles we obtained.

The last structure we analyzed were CHs. The ALMA and Mets\"ahovi measurements reveal that the CHs have slightly lower brightness temperatures than the measured QS (Table~\ref{Table_2}) and the ATLCW QS model (Figure \ref{Fig_5}). All output CH brightness temperature profiles indicate very little or almost no difference from the ATLCW QS level over the entire ALMA wavelength range, resulting in the CH being slightly darker than the QS at ALMA wavelengths. Moreover, our results also provide clear evidence that the brightness temperature of the CHs depends on their electron density rather than their temperature, which is consistent with the results of earlier observational and modeling studies by Braj{\v s}a et al. (\citeyear{Brajsa2007}, \citeyear{Brajsa2018a}).

Considering the CH density, out of three cases, the CH1\_best has significantly lower density when compared with the QS. Because the CH is at the solar limb, where more higher atmospheric layers are observed on average, one would expect to see a higher contrast in density, as well as temperature between a CH and the surrounding QS. The CH1\_best results for both the electron density and temperature could indicate just that. For a similar temperature, as is the situation with the three cases (CH1\_best, CH2\_best, and CH3\_best) individually, a lower density means that the CH profile shifts downward toward lower brightness temperatures. This is true for all three cases. However, having both lower density and temperature, one would expect a profile to have a larger downward shift, but all cases give similar profiles despite different output densities and temperatures. This could indicate that the chromospheric appearance of a CH visible in the ALMA wavelength range is not too sensitive to the density changes. Similarly, temperature changes as those seen among the three cases also do not affect the chromospheric appearance of a CH significantly. If there is any significant difference between the investigated cases, it might be at the coronal heights, outside the observed height range of the current measurements. To single out the more physical case for the given CH, we would first need to get measurements at even more wavelengths, which would increase the number of data points used in the profile fitting. Also, the current observed wavelength range might need to be extended to even coronal heights to constrain more the possible density and temperature parameters, for which the nonlinear Equation (\ref{Eq_radiative_transfer}) gives the best-fitting profile (future work).

Using the He I observations from the Solar tower telescope BST-2 and H$\alpha$ observations from the Global Oscillations Network Group (GONG) of a polar CH in the solar chromosphere, Baranovskii et al. (\citeyear{Baranovskii2019}) found the temperature of CHs to be in the range of 4\;580$-$8\;150 K, which is 500$-$1\;500 K lower than the QS level. This results in a CH temperature about 1.12$-$1.23 times lower than that of the surrounding QS. Also, based on the polar and equatorial CHs, Gopasyuk et al. (\citeyear{Gopasyuk2020}) found the CH temperatures to be 1.29$-$1.33 times lower than the QS value. The results of these two studies agree well with our results, especially with CH2\_best in Table \ref{Table_3}. In addition, Baranovskii et al. (\citeyear{Baranovskii2019}) found that the density of observed CHs in the solar chromosphere is 2$-$3 times lower than that of QS. Our results (Table \ref{Table_3}) give a CH density 1$-$1.7 times lower than the QS value, being below the minimum value found by Baranovskii et al. (\citeyear{Baranovskii2019}). When the uncertainties in the density factors are taken into account, the upper limit of our results falls inside the value range determined by these authors. This small difference between our results and those of Baranovskii et al. (\citeyear{Baranovskii2019}) could be because different CHs were observed, but also due to different spatial resolutions in the two studies.

It should be noted that the ATLCW, AL, and similar atmosphere models are hydrostatic models. This means that, for example, a modified temperature would change the pressure scale height, and thus the density profile. The obtained brightness temperature profiles for the observed AR and CH appear not to follow this trend because the profiles for both solar structures are less sensitive to the temperature changes and more sensitive to the density changes. From one aspect, when observing the brightness temperature of the solar atmosphere, more than one atmospheric layer in the observer's line-of-sight contributes to the overall measured brightness temperature. In the case of AR and CH, this means that a non-negligible change in density in the observed column of plasma in the solar atmosphere would cause a significant change in the observed brightness temperature. The change, significant or not, in only temperature in the observed plasma column by itself might not be enough to make as significant change in the measured intensity as the change in density would have. From another aspect, having different sensitivities on the density and temperature changes, the AR and CH structures as a whole might not be governed by hydrostatic laws, but rather some other laws or mechanisms, most likely related to magnetic field structure and its properties. If in fact the hydrostatic case is valid for AR and CH structures, it would not be applicable for a structure as a whole, but instead only within specific layers of a given structure.

Furthermore, we compare the current modeling results based on the modified ATLCW model presented in this work with earlier efforts by Braj{\v s}a et al. (\citeyear{Brajsa2018a}) based on the modified FAL model: 

(i) The QS brightness temperature profiles based on the ATLCW models are systematically higher than for the modified FAL models. The functional dependence of brightness temperature on wavelength is also different for the two models. 

(ii) The QS brightness temperature as a function of wavelength based on the ATLCW solar atmosphere model agrees much better with observations, both from ALMA, Mets\"ahovi, and from other studies, than the models presented by Braj{\v s}a et al. (\citeyear{Brajsa2018a}).

(iii) The models of ARs in the present work are more realistic and closer to the observational results than in the previous study (Braj{\v s}a et al. \citeyear{Brajsa2018a}).

(iv) The results for CHs are consistent at the qualitative level (no obvious difference from the QS level for wavelengths 0.3$-$4 mm and slightly lower emission of CHs for wavelengths 4$-$10 mm), but the functional dependence of brightness temperature on wavelength is different compared with the previous models. 

It should be noted that for structures other than QS we had only three observed wavelengths on which to base our analysis and conclusions. Therefore, more observations, especially above the wavelength of 3 mm, are needed to improve the current models, obtain more precise values of the density and temperature factors with lower uncertainties than the current results, and get a better picture of the physical properties of the solar structures. Moreover, depending on the spatial resolution (beam size), single-dish measurements could potentially lead to incorrect conclusions about the solar structure depending on how much of the surrounding region is included in the average brightness temperature measurement. For this reason, a comparison between the results of single-dish (as in this work) and interferometric observations could be useful to see if a model based on one type of observation is also valid for the other type (future work).

A similar analysis, as presented in this work, can also be applied to other solar structures such as prominences and sunspots. However, we should note that both prominences and sunspots are not background structures as QS, ARs, and CHs are. Prominences have complex three-dimensional structures that extend from the chromosphere to the corona, while sunspots are relatively small structures in the low solar atmosphere and single-dish observations are often not able to resolve them. This represents a challenge when modeling these two solar structures, so they should be treated differently than the solar structures analyzed in the present article. In future work, we plan to analyze both prominences and sunspots using a similar modeling as in this work, but including special treatments in order to study their complex two-dimensional and three-dimensional structures and properties in more details. 

The results of this work and their comparison with earlier studies suggest that the approach of using one-dimensional models based on the semi-empirical ATLCW model proved to be a good basis for modeling solar structures and sufficient for obtaining realistic brightness temperature values of those structures. We plan to extend the current analysis of this work by testing this approach for other solar structures not yet addressed in this work. However, the one-dimensional approach has a significant disadvantage compared to multidimensional models. For a given wavelength, the one-dimensional models only give the mean value of the brightness temperature of a given solar structure as a whole, without any details about the substructure. The next step would therefore be to extend the current one-dimensional model to a two-dimensional model, and later three-dimensional model. Using the procedure described in this work, one can further extend this procedure to obtain a two-dimensional brightness temperature map, together with corresponding electron density and temperature maps, of a given solar structure for different observing wavelengths (heights). We would then be able to study the cross-section of a solar structure and the variation of brightness temperature, and other physical properties within the structure. This will be addressed in future work.

\section*{Acknowledgments}

This work was supported by the Croatian Science Foundation as part of the "Young Researchers' Career Development Project - Training New Doctorial Students" under the project 7549 "Millimeter and submillimeter observations of the solar chromosphere with ALMA". Support from the Austrian-Croatian Bilateral Scientific Projects ”Comparison of ALMA observations with MHD-simulations of coronal waves interacting with coronal holes” and ”Multi-Wavelength Analysis of Solar Rotation Profile” is also acknowledged. It has also received funding from the Horizon 2020 project SOLARNET (824135, 2019–2023). In this paper, ALMA data ADS/JAO.ALMA\#2011.0.00020.SV were used. ALMA is a partnership of ESO (representing its member states), NSF (USA), and NINS (Japan), together with NRC (Canada), MOST and ASIAA (Taiwan), and KASI (Republic of Korea), in cooperation with the Republic of Chile. The Joint ALMA Observatory is operated by ESO, AUI/NRAO, and NAOJ. We thank the ALMA project for enabling solar observations with ALMA. This publication also uses data from the Mets\"ahovi Radio Observatory, operated by the Aalto University (Aalto University \citeyear{Aalto2019}). RB acknowledges financial support from the Alexander von Humboldt Foundation. HGL acknowledges financial support from the Deutsche Forschungsgemeinschaft (DFG, German Research Foundation)$-$Project-ID 138713538$-$SFB 881 ("The Milky Way System'', subproject A04). CLS acknowledges financial support from the S{\~a}o Paulo Research Foundation (FAPESP), grant number 2019/03301-8. We would also like to thank Manuela Temmer for critical reading of the manuscript.

\appendix
\bibliography{ASNA.20230149}


\end{document}